\documentclass[aps,prd,twocolumn,superscriptaddress,showpacs,floatfix,handout,10pt]{revtex4-1}

\usepackage{CJKutf8} %% to support Chinese, Japnese and Korean

\usepackage{amsmath,amssymb,mathtools}
\usepackage{color}
\usepackage{graphicx}
\usepackage{subfigure}
\usepackage[colorlinks=true,linkcolor=blue, citecolor=red, urlcolor=blue, bookmarks]{hyperref}
\usepackage{multirow,makecell} %% multirow and multicolumn
\usepackage{textcomp}
\usepackage{wasysym}
\usepackage{ulem}

\newcommand{\RNum}[1]{\uppercase\expandafter{\romannumeral #1\relax}}

\def \be {\begin{equation}}
\def \ee {\end{equation}}
\def \ba {\begin{array}}
\def \ea {\end{array}}
\def \bea {\begin{eqnarray}}
\def \eea {\end{eqnarray}}

\def \ble {\begin{widetext}\begin{equation}}
\def \ele {\end{equation}\end{widetext}}
\def \blea {\begin{widetext}\begin{eqnarray}}
\def \elea {\end{eqnarray}\end{widetext}}

\def \a {\alpha}

\def \D {\Delta}

\def \r {\rho}

\def \cO {{\mathcal O}}

\def \f {\frac}

\def \lag {\langle}
\def \rag {\rangle}

\def \dd {\mathrm{d}}
\def \ep {\mathrm{e}}
\def \ii {\mathrm{i}}

\def \tr {\mathrm{tr}}

\def \and {{\mathrm{and}}}

\def \ini {\textrm{ini}}

\def \tot {\textrm{tot}}

\def \ss {\textrm{ss}}

\begin{document}

\begin{CJK*}{UTF8}{gbsn}

\title{Subsystem Evolution Speed as Indicator of Relaxation}

\author{Jiaju Zhang (张甲举)}
\email{jiajuzhang@tju.eud.cn}
\affiliation{Center for Joint Quantum Studies and Department of Physics, School of Science,
Tianjin University, 135 Yaguan Road, Tianjin 300350, China}

\author{M.~A.~Rajabpour}
\email{mohammadali.rajabpour@gmail.com}
\affiliation{Instituto de Fisica, Universidade Federal Fluminense,
Av.~Gal.~Milton Tavares de Souza s/n, Gragoat\'a, 24210-346, Niter\'oi, RJ, Brazil}

\author{Markus Heyl}
\email{markus.heyl@uni-a.de}
\affiliation{Theoretical Physics III, Center for Electronic Correlations and Magnetism,
Institute of Physics, University of Augsburg, D-86135 Augsburg, Germany}

\author{Reyhaneh Khasseh}
\email{reyhaneh.khasseh@uni-a.de}
\affiliation{Theoretical Physics III, Center for Electronic Correlations and Magnetism,
Institute of Physics, University of Augsburg, D-86135 Augsburg, Germany}

\begin{abstract}

  For dynamical quantum matter many key features and properties address their steady states, such as the fundamental question of thermalization. However, it is a very concrete challenge to determine practically if a system has already relaxed and reached a steady state, as this requires a priori understanding of steady-state properties, which must, in principle, be verified across a large set of observables individually. In this paper, we introduce a general method to quantify the relaxation of a quantum state by measuring the speed of subsystem evolution. This approach allows us to assess the relaxation of all conceivable observables, providing a general measure of relaxation without requiring any a priori knowledge.

 % In studying the time evolution of isolated many-body quantum systems, a key focus is determining whether the system undergoes relaxation and reaches a steady state at a given point in time. Traditional approaches often rely on specific local operators or a detailed understanding of the stationary state. In this paper, we introduce an alternative method that assesses relaxation directly from the time-dependent state by focusing on the evolution speed of the subsystem. The proposed indicator evaluates the rate of change in the reduced density matrix of the subsystem over time. We demonstrate that in systems reaching relaxation, as the overall system size increases, the evolution speed of sufficiently small yet still finite-sized subsystems notably diminishes. This leads to small fluctuations in the expectation values of operators, which is also consistent with the predictions made by the eigenstate thermalization hypothesis. We apply this approach across various models, including the chaotic Ising chain, XXZ chains with and without many-body localization, and the transverse field Ising chain. Our results confirm the robustness and accuracy of subsystem evolution speed as a reliable indicator for relaxation.

\end{abstract}

\maketitle

\end{CJK*}

%\tableofcontents

%\section{Introduction}

\section{Introduction}
Relaxation stands as a pivotal concept in the study of quantum dynamics \cite{Yukalov:2011tvz,Ueda:2020cve}. It describes how a quantum system evolves towards a steady state that remains invariant over time. A critical inquiry within this domain concerns the mechanisms that govern thermalization, a specific form of relaxation where the system reaches thermal equilibrium, and observables align with the predictions of statistical mechanics. Depending on the system's inherent dynamics, chaotic systems typically thermalize to a Gibbs ensemble, fully erasing the memory of their initial conditions, whereas integrable systems gravitate towards a generalized Gibbs ensemble (GGE), retaining some information about their initial state. This distinction underscores the diverse pathways to relaxation, shaped by whether the system is chaotic or integrable \cite{Rigol:2006jrd,Mandal:2015jla,Cardy:2015xaa,Mandal:2015kxi,Vidmar:2016laa,deBoer:2016bov}.

In 1929, von Neumann pioneered the discussion on how quantum systems under unitary dynamics could mirror statistical mechanics behavior \cite{vonNeuman:1929une}. He introduced the idea of thermalization from the perspective of observables, suggesting that an observable achieves thermalization when its long-term behavior closely matches the predictions of the microcanonical ensemble. The insights of von Neumann align well with the predictions of Random Matrix Theory (RMT) \cite{DAlessio:2015qtq,Borgonovi:2016hnp}. However, to accurately describe observables in experiments, one must extend beyond RMT, as real systems exhibit thermal expectation values influenced by their energy density (temperature), a nuance not captured by RMT. Moreover, the matrix elements of observables in real systems encode information beyond the scope of RMT predictions. The eigenstate thermalization hypothesis (ETH) addresses this gap, incorporating the nuances of thermalization in physical observables \cite{Deutsch:1991msp,Srednicki:1994mfb,Rigol:2007mja}

Two conventional approaches exist for assessing whether a quantum system has reached a steady state, with thermalization being a specific instance. The first method involves examining the expectation values of local operators within a subsystem \cite{Banuls:2010fkc}. If these expectation values remain unchanged over time, it indicates that the subsystem settled into a steady state. However, this approach requires the evaluation of all linearly independent operators within the Hilbert space of the subsystem, posing an intricate challenge, while the inclusion of higher-order observables also remains an open problem. Alternatively, the second method assesses relaxation by calculating the distance between the subsystem reduced density matrix (RDM) and the RDM of the steady state \cite{Banuls:2010fkc,Fagotti:2013jzu,Dymarsky:2016ntg}. While this method provides a direct metric for relaxation, it necessitates prior knowledge of the exact steady state, which can be a challenging prerequisite to fulfill.

In the paper, we propose an alternative relaxation indicator that relies solely on the quantum system's state, eliminating the necessity for local operators or a predefined steady state. This indicator quantifies the rate of change of the time-dependent subsystem RDM, thereby representing the subsystem's evolution speed in a manner analogous to the geometric approach to the quantum speed limit \cite{Mandelstam:1945mns,Margolus:1997ih,Taddei:2013sbt,delCampo:2013koe,Deffner:2013uar,Frey:2016vil,Deffner:2017cxz}.

%\section{Definition of evolution speed}

\section{Definition of evolution speed}
To define evolution speed, we first need a measure of distance that effectively captures changes between quantum states. The concept of distance between two quantum states can be defined in several ways \cite{Nielsen:2010oan,Watrous:2018rgz}, with the trace distance often considered the most effective \cite{Fagotti:2013jzu,Zhang:2020mjv}.
The trace distance between the density matrices $\rho$ and $\sigma$ is given by
\begin{equation}
D(\rho, \sigma) = \frac{1}{2} \left\|\rho - \sigma\right\|_1,
\end{equation}
with $\|\cdot\|_1$ denoting the trace norm.
This formulation ensures that the trace distance always falls within the range $0 \leq D(\rho, \sigma) \leq 1$. Considering a state $\rho(t)$ that varies over time, which may represent the density matrix of an isolated quantum system or the RDM of a subsystem, we define the evolution speed of the system as
\begin{equation}
v(t) \equiv \lim_{\delta t \to 0} \frac{D(\rho(t), \rho(t + \delta t))}{\delta t},
\end{equation}
which effectively quantifies the rate at which the state $\rho(t)$ changes over time.

For an isolated quantum system governed by the Hamiltonian $H(t)$ with a pure state $|\psi(t)\rangle$
%\begin{equation}
%|\psi(t)\rangle = \cT \exp \left( -i \int_0^t H(\tau) d\tau \right) |\psi(0)\rangle,
%\end{equation}
%with the time-ordering operator $\cT$ and the initial state $|\psi(0)\rangle$.
the evolution speed, which measures how quickly the total quantum state changes, is given by
\begin{equation}
v_\tot (t) = \lim_{\delta t \to 0} \frac{D(|\psi(t)\rangle, |\psi(t + \delta t)\rangle)}{\delta t}.
\end{equation}
Using the trace distance between two general pure states $D(|\psi\rangle, |\phi\rangle) =
\sqrt{1 - |\langle\psi|\phi\rangle|^2}$, we get the evolution speed of the total system as the fluctuation of the energy
\begin{equation}
\begin{split}
v_{\tot} (t) & = \Delta H(t) \\
            & \equiv \sqrt{\langle\psi(t)|[H(t)]^2 |\psi(t)\rangle - \langle\psi(t)|H(t)|\psi(t)\rangle^2},
\end{split}
\end{equation}
which is consistent with the Mandelstam-Tamm bound of the quantum speed limit \cite{Mandelstam:1945mns}.
When the Hamiltonian is time-independent, the total system evolves with a constant speed $v_{\tot} = \Delta H$.

For subsystems, however, the analysis is not as straightforward due to additional complexities arising from mixed state dynamics. For a subsystem $A$, the time-dependent RDM can be obtained by
\begin{equation}
\rho_A (t) = \mathrm{tr}_{\bar{A}} |\psi(t)\rangle\langle\psi(t)|.
\end{equation}
with $\bar{A}$ denoting the complementary subsystem, and the evolution speed can be defined as
\begin{equation}\label{eq:speed_sub}
v_A (t) = \lim_{\delta t \to 0} \frac{D(\rho_A (t), \rho_A (t + \delta t))}{\delta t}.
\end{equation}
From the contractive property of the trace distance under partial trace \cite{Nielsen:2010oan,Watrous:2018rgz}
\begin{equation}
D(\rho_A (t), \rho_A (t + \Delta t)) \leq D(|\psi(t)\rangle, |\psi(t + \Delta t)\rangle),
\end{equation}
we get
\begin{equation}
v_A (t) \leq v_\tot (t) = \Delta H(t).
\end{equation}
As long as the fluctuation of the energy $\Delta H(t)$ is finite, the evolution speed for the total system $v_{\tot} (t)$ and the subsystem evolution speed $v_A (t)$ are well-defined
with no divergence in the $\delta t \to 0$ limit.

Alternative definitions of distance measures and the corresponding evolution speeds are explored and discussed in Sec.~\ref{supp:s2} of  appendix.

%\section{Indicators of relaxation}

\section{Indicators of relaxation}
The relaxation process of a time-varying pure state, represented as $|\psi(t)\rangle$, typically progresses towards a steady state, denoted by $\rho_{\textrm\ss}$.
%This process can be examined through several approaches. One straightforward method involves calculating the expected values of local operators within the given state. As relaxation occurs, the expected values within a subsystem tend to stabilize. In cases where thermalization takes place, this stabilization reflects the ETH, with observable values averaging out to match those predicted by the thermal state. When employing this method, it may be necessary to consider all linearly independent operators within the subsystem's Hilbert space, which can be quite challenging. %Additionally, the specific operators required may vary depending on the particular time-dependent state $|\psi(t)\rangle$ being analyzed.
%
%Alternatively, relaxation can be assessed by computing the distance $D(\rho_A (t), \rho_{A,\text\ss})$ between the time-dependent RDM $\rho_A (t) = \text{tr}_{\bar{A}} |\psi(t)\rangle\langle\psi(t)|$ and the RDM of the steady state $\rho_{A,ss} = \text{tr}_{\bar{A}} \rho_\ss$. This distance quantifies the deviation between the state of the subsystem and its steady state, providing a direct measure of relaxation \cite{Banuls:2010fkc,Fagotti:2013jzu,Dymarsky:2016ntg}.
%
In the study of a quantum spin chain of $L$ total sites and a subsystem with $L_A$ consecutive sites, significant interest lies in exploring the scaling limit where $L \to +\infty$ and $L_A \to +\infty$, maintaining a constant ratio $x = L_A/L$. Upon reaching relaxation, either as $t$ approaches infinity or at some specific time $t = t_*$, a critical ratio $x_*$ emerges.
For subsystem sizes below the threshold $x < x_*$, the trace distance between the time-dependent RDM $\rho_A (t)$ and its steady state counterpart $\rho_{A,\text\ss}$ approaches zero
\begin{equation}
D_A^\ss(t) \equiv D(\rho_A (t), \rho_{A,\ss}) \to 0,
\end{equation}
indicating that relaxation has occurred. Conversely, due to the monotonic decrease of the trace distance under partial trace \cite{Nielsen:2010oan,Watrous:2018rgz}, as the ratio $x$ goes from $x_*$ to 1, the subsystem trace distance $D_A^{f}(t)$ increases monotonically from 0 to the total system distance
\begin{equation}
D_\tot^\ss(t) \equiv D(|\psi(t)\rangle, \rho_\ss).
\end{equation}
%reflecting the increasing influence of the total system's state on the subsystem.

Under a time-independent Hamiltonian, the total system distance $D_\tot^\ss(t)$ remains constant.
In this scenario, pinpointing the exact steady state $\rho_\ss$ to which the pure state $|\psi(t)\rangle$ evolves becomes crucial. For chaotic systems, this state typically corresponds to the thermal state described by the Gibbs ensemble, whereas integrable systems tend to relax to a GGE state \cite{Rigol:2006jrd}. Generally, the task of defining or identifying $\rho_\ss$ for these systems can be elusive and challenging.

To address this challenge, we leverage the subsystem evolution speed $v_A(t)$ (\ref{eq:speed_sub}) as an indicator of relaxation, thereby bypassing the complexities associated with finding the proper local operators or directly determining the steady state.
As a subsystem approaches a steady state, its evolution speed $v_A(t)$ progressively decreases in the thermodynamic limit and ultimately approaches zero. This behavior occurs when the ratio $x=L_A/L$ is less than a critical value, typically $x_* = 1/2$. For a finite system size $L$, the subsystem evolution speed $v_A(t)$ remains finite, but this value decreases as the size of the system increases, indicating the system's progression towards a steady state.

When the steady state is reached at the infinite time, fluctuations in the subsystem evolution speed $v_A(t)$ may occur at some large but finite time. To smooth out these fluctuations, we take the time average of the evolution speed over a sufficiently long period, denoted as $\langle v_A(t) \rangle$. Since the total system evolution speed $v_\tot$ generally increases with the size of the system, we normalize the average subsystem evolution speed $\langle v_A(t) \rangle$ by $v_\tot$ as $\langle v_A(t) \rangle/v_\tot$.

When the steady state is reached at a finite time, say $t=t_*$, time averaging is unnecessary.
%However, normalization by $v_\tot$ is still required, so we consider $\langle v_A(t_*) \rangle/v_\tot$.
Specific examples of these cases are discussed further in Sec.~\ref{supp:s3} of the  appendix.

In Sec.~\ref{supp:s4} and Sec.~\ref{supp:s5} of the  appendix, we show that an exponential decrease in the subsystem evolution speed as a function of system size leads to exponentially small time fluctuations in the operator's expectation value. This behavior is consistent with the predictions of the ETH, indicating that systems displaying such characteristics in evolution speed achieve relaxation. We also show that exponential small subsystem trace distance leads to exponential small subsystem evolution speed.

%\section{Numerical checks}

\section{Numerical checks}
We investigate the dynamics of the evolution speed metric within a range of models, beginning with the Ising chain subjected to both transverse and longitudinal fields, which is anticipated to exhibit thermalization to the canonical ensemble \cite{Banuls:2010fkc,Kim:2015cdh}, with $\rho_\ss = \rho_{\text{thermal}}$.
%Subsequently, we delve into the transverse field Ising model, wherein the resulting equilibrium state aligns with the GGE state \cite{Calabrese:2011vdk,Cazalilla:2011ivg,Calabrese:2012fmm,Calabrese:2012lgi}, denoted as $\rho_\ss = \rho_{\text{GGE}}$.
We also analyze the spin-$\f12$ XXZ chain subjected to random disorder \cite{Pal:2010ezp,Luitz:2015cep}, which exhibits characteristics of integrable, ETH and many-body localized (MBL) Hamiltonians \cite{Alet:2018dwx,Sierant:2024khi}.

%\subsection{Chaotic Ising chain}

\subsection{Chaotic Ising chain}
We first study the chaotic Ising chain, characterized by the Hamiltonian
\begin{equation}\label{eq:Chaotic_Ising}
H(h_x, h_z) = -\frac{1}{2} \sum_{j=1}^{L} (\sigma_j^x \sigma_{j+1}^x + h_x\sigma_j^x + h_z\sigma_j^z).
\end{equation}
This system features both a longitudinal field $h_x$ and a transverse field $h_z$, with the Hamiltonian exhibiting chaotic behavior provided all tunable parameters are nonzero. Although the model is expected to reach a thermal state over a long period, this is not the case for all initial states in which the system is initialized. In the following, we present the behavior of the evolution speed of the subsystem for different initial states and compare these results with previously established metrics to verify their consistency.

\begin{figure}[ht]
  \includegraphics[height=0.22\textwidth]{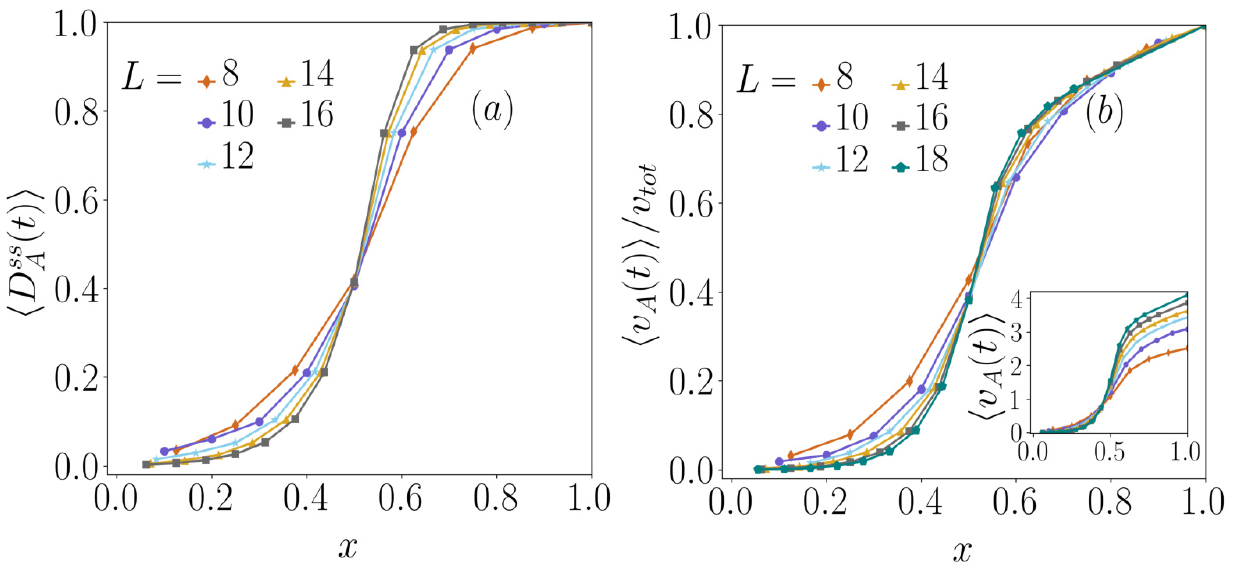}\\
  \caption{%
  {\bf Chaotic Ising chain}: Subsystem trace distance and subsystem evolution speed with an initial random state.
  %(a) Dynamics of the trace distance $D_A^\ini(t)$ between the RDM at time $t$ and the RDM at the initial time.
  %(b) Dynamics of the trace distance $D_A^\ss(t)$ between the RDM at time $t$ and its corresponding steady state RDM.
  %(c) Dynamics of the subsystem evolution speed $v_A(t)$.
  (a) Long-time average of $D_A^\ss(t)$ as a function of the ratio $x = \f{L_A}{L}$.
  (b) Long-time average of $v_A(t)$ relative to the total system evolution speed $v_\tot$. The inset shows the subsystem evolution speed without normalization.
  The parameters are $h_x = \frac{\sqrt{3}}{2}$ and $h_z = \sqrt{2}$.
  %For panels (a), (b), and (c), a total system size of $L = 16$ is used.
  The time averages for the distance and speed are taken over the interval $t \in [L, 2L]$.%
  }
  \label{FigureChaoticIsingRandom}
\end{figure}

We begin with an initial state $|\psi(0)\rangle$ that is a random state, with coefficients drawn from a Gaussian distribution. Such random states typically have significant overlap with all energy eigenstates, making them well-suited for observing relaxation and thermalization. The chaotic Ising model equilibrates to the thermal state, represented by a maximally mixed state $\rho_\ss = 1/2^L$.
%In Fig.~\ref{FigureChaoticIsingRandom}(a), (b), and (c), we illustrate the evolutions of the trace distance $D_A^\text{ini}(t)$ between the time-dependent RDM $\rho_A(t)$ and the initial RDM $\rho_A(0)$, the trace distance $D_A^\ss(t)$ between the time-dependent RDM $\rho_A(t)$ and its equilibrium counterpart $\rho_A^\ss$, and the subsystem evolution speed $v_A(t)$, across various subsystem sizes. All plots demonstrate the emergence of stationary behavior over time.
In Fig.~\ref{FigureChaoticIsingRandom}, panels (a) and (b) show the time-averaged values of the distance and velocity metrics, respectively, in the steady state as functions of the subsystem size ratio $x = L_A/L$.
Notably, for subsystems smaller than half the size of the total system, the subsystem evolution speed $v_A(t)$, normalized by the total system evolution speed $v_\tot$, follows similar dynamics to that of the subsystem trace distance $D_A^\ss(t)$.
For fixed $0 < x < 1/2$, the trace distance between the RDM and the corresponding thermal RDM decreases with increasing system size, suggesting convergence towards the relaxation state in the thermodynamic limit. Similarly, the observed decrease in evolution speed with increasing total system size further supports the trend toward relaxation.

\begin{figure}[ht]
 \includegraphics[height=0.44\textwidth]{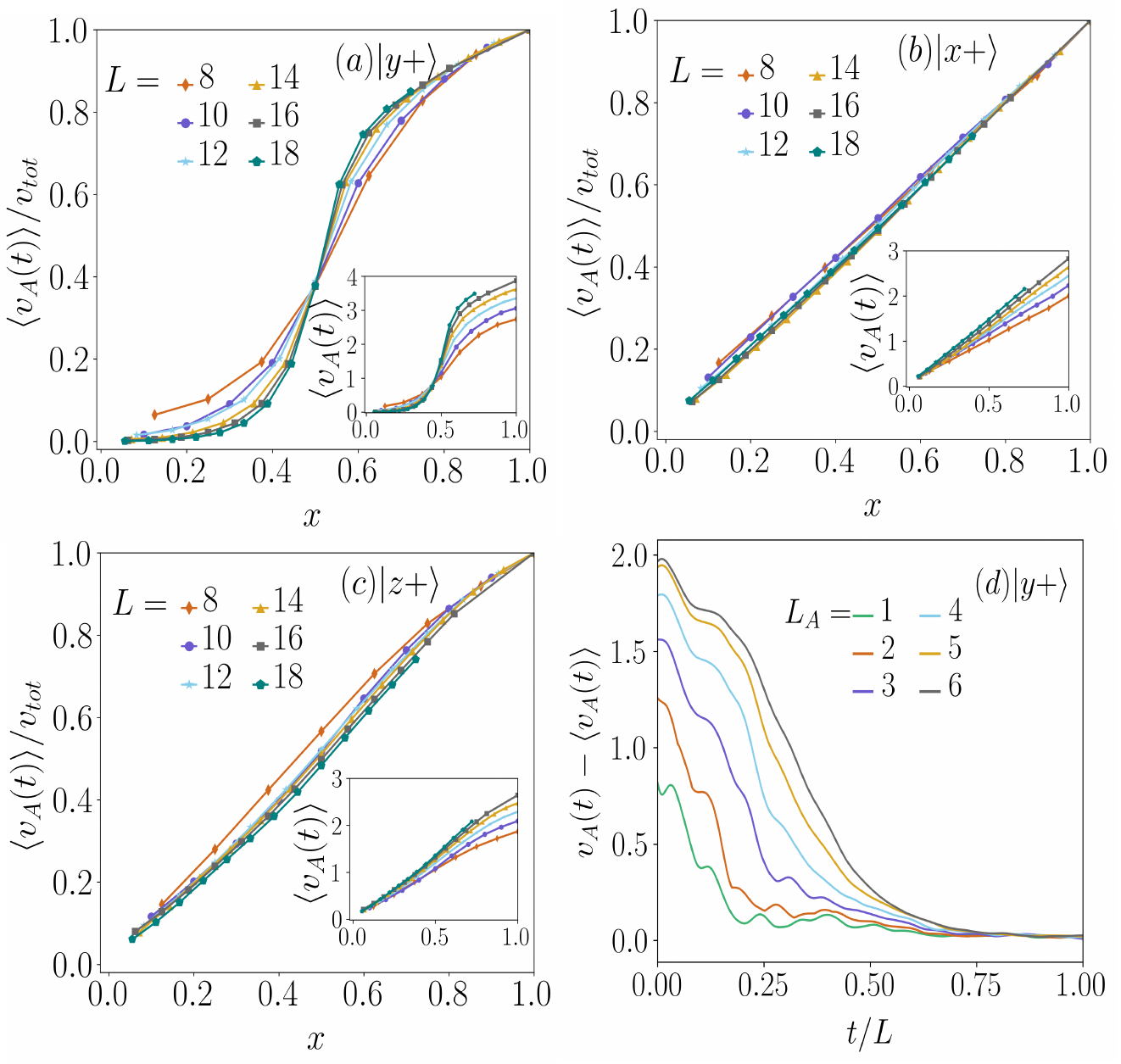}\\
  \caption{
  {\bf Chaotic Ising chain}: (a), (b), and (c) show the long-time average of the velocity as a function of $x$ for the initial states $|y+\rangle$, $|x+\rangle$, and $|z+\rangle$, respectively. 
  The insets display the corresponding subsystem evolution speed without normalization. (d)  Dynamics of the deviation of the subsystem evolution speed from its time-averaged value for the initial state $|y+\rangle$, and $L=16$.
  We have used $h_x = \frac{\sqrt{3}}{2}$ and $h_z = \sqrt{2}$. 
  The time averages are taken over $t \in [L, 2L]$.% 
}
  %\caption{%
  %{\bf Chaotic Ising chain}: Long-time average of the subsystem evolution speed over the total system evolution speed with the initial state being a product state:
  %(a) $|x+\rangle$, (b) $|y+\rangle$, (c) $|z+\rangle$, and (d) $|\text{N\'{e}el}\rangle$.
  %The insets show the subsystem evolution speed without normalization.
  %We have used $h_x = \frac{\sqrt{3}}{2}$ and $h_z = \sqrt{2}$.
  %The time averages are over $t \in [L, 2L]$.%
  %}
\label{FigureChaoticIsingProduct}
\end{figure}

As a second case, we use initial states that are direct product states, such as %
$%
|x+\rangle \equiv \bigotimes_{j=1}^{L} \frac{| \uparrow_j \rangle + | \downarrow_j \rangle}{\sqrt{2}}, ~
|y+\rangle \equiv \bigotimes_{j=1}^{L} \frac{| \uparrow_j \rangle + \ii | \downarrow_j \rangle}{\sqrt{2}}, ~
|z+\rangle \equiv \bigotimes_{j=1}^{L} | \uparrow_j \rangle, ~
|\text{N\'{e}el}\rangle \equiv |\uparrow\downarrow\uparrow\downarrow\cdots\rangle%
$. %
The dynamics of the chaotic Ising model initialized in a product state, have been extensively analyzed in \cite{Banuls:2010fkc}, where two distinct thermalization regimes were identified. The state $|y+\rangle$ exhibits strong thermalization, with its instantaneous expectation values rapidly converging to those of the thermal state. In contrast, the state $|x+\rangle$ demonstrates weak thermalization, where the state does not relax even at very long times, although the time-averaged expectation values of observables eventually reach thermal values. In Fig.~\ref{FigureChaoticIsingProduct}(a) and (b), the average evolution speeds corresponding to both strong and weak thermalization regimes are depicted. For the initial state $|y+\rangle$, the average evolution speed $\lag v_A(t) \rag$ decays towards zero for $x<1/2$, suggesting convergence to relaxation, which in this context is the thermal state. Conversely, for the state $|x+\rangle$, there is no indication of relaxation and weak thermalization.
When the system is initialized in $|z+\rangle$, it shows no signs of relaxation over a long time; see Fig.~\ref{FigureChaoticIsingProduct}(c).

In Fig.~\ref{FigureChaoticIsingProduct}(d) we further investigate another interesting application of the subsystem evolution speed, concerning the temporal relaxation as a function of subsystem size. While the relaxation of local observables by now is well understood, this is not the case for nonlocal operators such as long Pauli strings. A systematic analysis of such relaxation is overall hindered as a complete brute-force analysis would require the study of a large number of such Pauli string, which grows exponentially with the string length. This is not only of relevance in order to understand theoretically such relaxation, but it also becomes important in experiments~\cite{Bandyopadhyay:2021,Karch:2025}.

Setting $h_x=0$ in (\ref{eq:Chaotic_Ising}) transitions the system to the transverse field Ising model (TFIM), which is integrable and can be solved analytically \cite{Lieb:1961fr, Pfeuty:1970ayt}. Over a long period, this system can be relaxed to the GGE state \cite{Calabrese:2011vdk, Cazalilla:2011ivg, Calabrese:2012fmm, Calabrese:2012lgi}. %, denoted as $\rho_\ss = \rho_{\text{GGE}}$.
In Sec.~\ref{supp:s3} of the supplemental material, we consider examples of the initial random state and product states within the TFIM. We also examine the time evolution of subsystems after a quantum quench \cite{Calabrese:2005in, Calabrese:2006rx, Cardy:2014rqa, Alba:2017hlc}, for which the Bures distance and the corresponding evolution speed can be efficiently calculated following \cite{Banchi:2013uht}.

%\subsection{XXZ chains outside of and in MBL regime}

\subsection{XXZ chains outside of and in MBL regime.}
The Hamiltonian for the XXZ model is given by
\begin{equation}\label{eq:Hamiltonian_XXZ}
\hspace{-0.8mm}H = \sum_{j=1}^{L} \Big[ \frac{1}{4} \big( \sigma_j^x\sigma_{j+1}^x + \sigma_j^y\sigma_{j+1}^y + \Delta\sigma_j^z\sigma_{j+1}^z \big)
                             + \frac{1}{2} h_z^j\sigma_j^z \Big],
\end{equation}
where $\Delta$ represents the anisotropy parameter.
We focus on a scenario where $h_z^j$ varies randomly within the range $[-h, h]$, adhering to a uniform distribution. This random variation introduces disorder, which is crucial for modeling many-body localization (MBL) and allows us to investigate its impact on the dynamics of the quantum many-body system \cite{Pal:2010ezp,Luitz:2015cep}.
When the disorder parameter $h=0$, the system reduces to the integrable XXZ chain \cite{Caux:2014uuq,Franchini:2016cxs}.
As $h$ increases, the random field XXZ chain goes from the ETH region to the MBL region.

%For the magnetic field $h_z^j$, we consider a uniform field scenario, $h_z^j = h_z$, which simplifies the Hamiltonian to the integrable XXZ model. The analysis and results for this case are presented in the appendix.

The MBL phase is characterized by the absence of thermalization in isolated many-body systems, influenced by both disorder and interactions, though the system may relax to a non-thermal steady state. Crucially, this characterization requires considering the limits of both infinitely long-time evolution and infinite system size. Definitive conclusions about the MBL phase are constrained by the small system sizes typically used in numerical studies. Over a decade of research into the MBL phase has left many questions and possibilities open. In this paper, we focus on the concept of relaxation within systems exhibiting MBL phases by analyzing the subsystem evolution speed. Despite the dynamics observed with integrable and non-integrable Hamiltonians, the relaxation state under MBL Hamiltonian evolution remains undefined, meaning there is no clear $\rho_\ss$ in this context. This ambiguity underscores the utility of alternative measures, such as the subsystem evolution speed, to predict and analyze relaxation dynamics.

\begin{figure}[ht]
  \includegraphics[height=0.44\textwidth]{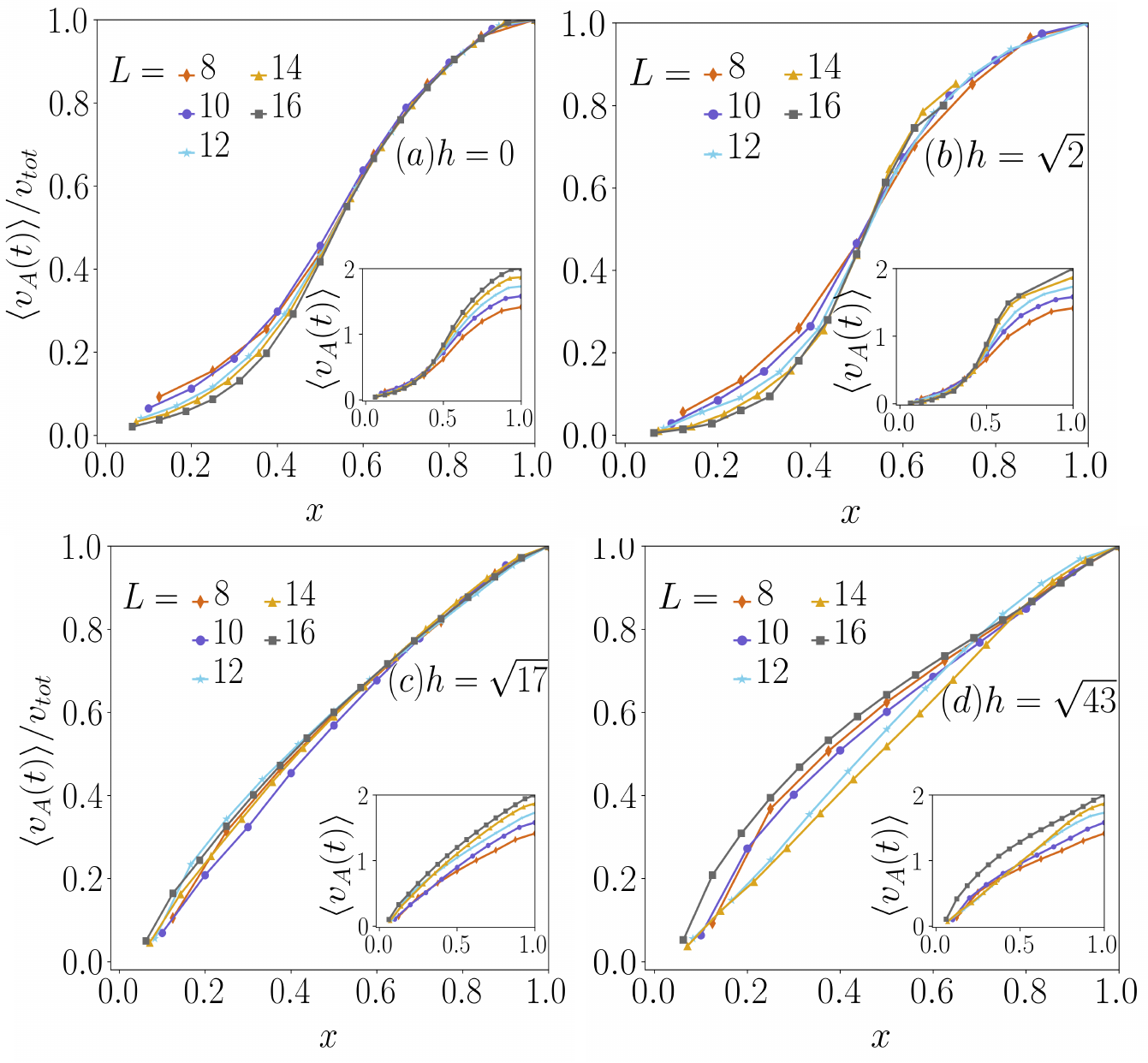}\\
  \caption{%
  {\bf XXZ chain}: Long-time average of the subsystem evolution speed with the initial state being the N\'{e}el state with different strengths of the disorder with $h=0,\sqrt{2},\sqrt{17},\sqrt{43}$ from left to right.
  The insets show the subsystem evolution speed without normalization.
  In all panels, we have used $\D = 1$.
  The time averages are over $t \in [4L, 8L]$.%
  }
\label{FigureXXZMBLNeel}
\end{figure}

A single Hamiltonian (\ref{eq:Hamiltonian_XXZ}), with random fields already chosen from a uniform distribution, is not strictly translationally invariant. Therefore, when calculating the subsystem distance and evolution speed we consider all $L$ possible starting positions of the subsystem and then take the average. In Fig.~\ref{FigureXXZMBLNeel}, we illustrate the dependency of the subsystem evolution speed on system size for the initial N\'{e}el state. In panels (a) and (b), the model is, respectively, integrable and in the ETH regime, and the N\'{e}el state reaches steady states in both cases. In panels (c) and (d), there is strong disorder, and the model is in the MBL regime; here, there is no sign of relaxation for the N\'{e}el state. This is expected, as the N\'{e}el state is approximately an energy eigenstate in this regime, and there is no ETH for the random field XXZ chain exhibiting MBL. Additional details and further analysis of the ETH–MBL transition can be found in Sec. \ref{supp:s1} of appendix.

%\section{Conclusions}

\section{Conclusions}
In this paper, we have proposed utilizing the subsystem evolution speed, which corresponds to the rate of change of the reduced density matrix of the subsystem, as an indicator of relaxation in quantum systems. This novel approach relies solely on the state of the quantum system, negating the need for specific local operators or prior knowledge of the steady state. As the size of the quantum system grows, the evolution speed of sufficiently small subsystems decreases, signaling the onset of relaxation. This method has been validated on various models, including chaotic and integrable systems, demonstrating its robustness and precision. Our approach is independent of numerical methods and system dimensionality, making it broadly applicable. However, efficient computation of system states and reduced density matrices remains challenging for larger systems and longer times. Extending to more complex quantum systems—higher dimensions, long-range interactions, or diverse symmetries—is a natural direction. For 1D long-range interactions, exact diagonalization (ED) remains viable, reaching system sizes similar to short-range cases. In 2D, tensor networks (PEPS, TTN) and hybrid methods (tDMRG, variational Monte Carlo) can simulate dynamics but are typically limited concerning the accessible time scales to short evolution times. %It would be interesting to extend it to more intricate quantum systems, such as those with higher dimensions, long-range interactions, and diverse symmetries. Such extensions would not only validate the robustness and generality of the method but also broaden its applicability to a wider range of quantum systems.

Furthermore, the subsystem evolution speed might enable to address a rather open question in quantum many-body dynamics, which concerns the relaxation of nonlocal operators. While the dynamics of local observables are by now well understood, it has remained largely unclear how on general grounds expectation values of nonlocal observables such as long Pauli strings relax, relevant also for recent experiments~\cite{Bandyopadhyay:2021,Karch:2025}. As the subsystem evolution speed characterizes the relaxation of the full reduced density matrix it automatically accounts for all nonlocal operators, which is otherwise all but straightforward as the number of all Pauli strings of length $l$, say, grows exponentially in $l$. Moreover, developing a deeper theoretical understanding of how subsystem evolution speed relates to key dynamical processes in quantum systems, such as relaxation and thermalization, is crucial. By exploring how factors like disorder, interactions, and dimensionality influence evolution speed, we can gain valuable insights into the fundamental mechanisms driving these processes. In particular, studying the behavior of evolution speed in systems exhibiting MBL and near critical points could offer critical insights into the transition from integrable to ergodic dynamics.

%\section*{Acknowledgements}

%{\bf Acknowledgements.}
\section*{Acknowledgements}
JZ thanks support from the National Natural Science Foundation of China (NSFC) grant number 12205217. MAR thanks CNPq and FAPERJ (Grant
No. E-26/210.062/2023) for partial support.
MH has received funding from the European Research Council (ERC) under the European Unions Horizon 2020 research and innovation program (grant agreement No. 853443). This work was supported by the German Research Foundation DFG via project 499180199 (FOR 5522). The authors gratefully acknowledge the resources on the LiCCA HPC cluster of the University of Augsburg, co-funded by the Deutsche Forschungsgemeinschaft (DFG, German Research Foundation)-Project-ID 499211671.

\hfill

%\newpage

%\providecommand{\href}[2]{#2}
\begingroup\raggedright\endgroup

%\bibliographystyle{D:/00.bibx/JHEPx}
%\bibliography{D:/00.bibx/2024,D:/00.bibx/2023,D:/00.bibx/2022,D:/00.bibx/2021,D:/00.bibx/2020,D:/00.bibx/2019,D:/00.bibx/2018,D:/00.bibx/1900,D:/00.bibx/1960,D:/00.bibx/1970,D:/00.bibx/1980,D:/00.bibx/1990,D:/00.bibx/1995,D:/00.bibx/1996,D:/00.bibx/1997,D:/00.bibx/1998,D:/00.bibx/1999,D:/00.bibx/2000,D:/00.bibx/2001,D:/00.bibx/2002,D:/00.bibx/2003,D:/00.bibx/2004,D:/00.bibx/2005,D:/00.bibx/2006,D:/00.bibx/2007,D:/00.bibx/2008,D:/00.bibx/2009,D:/00.bibx/2010,D:/00.bibx/2011,D:/00.bibx/2012,D:/00.bibx/2013,D:/00.bibx/2014,D:/00.bibx/2015,D:/00.bibx/2016,D:/00.bibx/2017,D:/00.bibx/book,D:/00.bibx/work,D:/00.bibx/thesis}

\clearpage

\newpage

%\onecolumngrid

%change the format of page, section, equation, figure and table%
%\renewcommand\thepage{S\arabic{page}}
%\renewcommand\thesection{S\arabic{section}}
%\renewcommand{\thesubsection}{S\arabic{subsection}}
%\renewcommand\theequation{S\arabic{equation}}
%\renewcommand\thefigure{S\arabic{figure}}
%\renewcommand\thetable{S\arabic{table}}

%\setcounter{page}{1}
%\setcounter{section}{0}
%\setcounter{equation}{0}
%\setcounter{figure}{0}
%\setcounter{table}{0}

\newpage

\appendix

%\begin{appendices}
\addtocontents{toc}{\protect\setcounter{tocdepth}{1}}
\renewcommand{\theequation}{\thesection.\arabic{equation}}

\setcounter{equation}{0}

\section{Transition to MBL phase in XXZ chain}\label{supp:s1}
\begin{figure*}[ht]
  \includegraphics[height=0.39\textwidth]{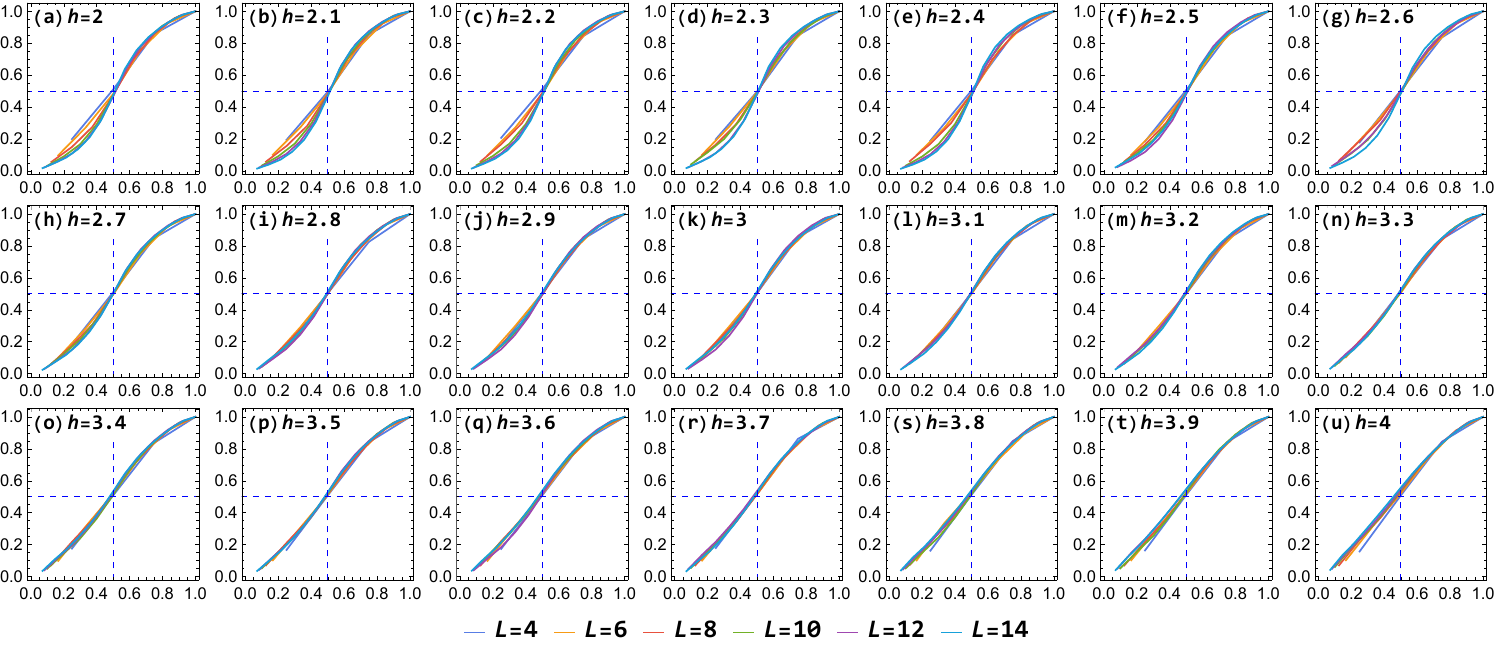}\\
  \caption{%
  {\bf XXZ chain}: The average subsystem evolution speed behaves near the transition from ETH to MBL. The horizontal axes is the
ratio $L_A/L$ and the vertical axes is the average subsystem evolution speed $\langle v_A\rangle/\langle v_T\rangle$. We have used $N_r = 32$ realizations of the
random fields.
 }
\label{MBL_transition}
\end{figure*}

To examine the behavior of the average subsystem evolution speed near the transition between the ergodic (ETH) and many-body localized (MBL) phases, we present additional results in Fig.~3. We focus on the parameter range 
\[
2 \leq h \leq 4,
\]
which spans both regimes and includes the critical region where the transition occurs (around \(h \approx 3\) \cite{Luitz:2015cep}). To minimize statistical fluctuations and better reveal the underlying trend, we conduct a three-fold averaging procedure, encompassing the time interval $t\in[4L,8L]$, all $L$ possible positions of subsystem $A$ along the circular chain, and 32 independent realizations of the random fields.
The resulting data in Fig.~\ref{MBL_transition} demonstrate that the average subsystem evolution speed undergoes a noticeable change in behavior in the vicinity of \(h \approx 3\), thereby providing compelling evidence for the transition from the ETH to the MBL regime.

\section{Alternative definitions of the evolution speed}\label{supp:s2}

In the main text, we have considered the trace distance as the distinguishably of two quantum states. Here, we consider alternative definitions of the distance and the corresponding evolution speed.

The Bures distance between the quantum states $\rho$ and $\sigma$ is defined as
\begin{equation} \label{Brhosigma}
B(\rho, \sigma) = \sqrt{2[1 - F (\rho, \sigma)]},
\end{equation}
where the fidelity $F (\rho, \sigma)$ is given by
\begin{equation}
F (\rho, \sigma) = \mathrm{tr} \sqrt{\sqrt{\rho} \sigma \sqrt{\rho}}.
\end{equation}
By definition, $0 \leq F (\rho, \sigma) \leq 1$, and hence, $0 \leq B(\rho, \sigma) \leq \sqrt{2}$.
The trace distance $D(\rho,\sigma)$ and fidelity $F(\rho,\sigma)$ are equivalent in the sense of the inequalities \cite{Nielsen:2010oan,Watrous:2018rgz}
\begin{equation}
\sqrt{1 - F (\rho, \sigma)} \leq D(\rho, \sigma) \leq \sqrt{1 - F (\rho, \sigma)^2},
\end{equation}
which further establishes an equivalence between the trace distance $D(\rho,\sigma)$ and Bures distance $B(\rho,\sigma)$ through the inequalities
\begin{equation}
\frac{B(\rho, \sigma)^2}{2} \leq D(\rho, \sigma) \leq B(\rho, \sigma) \sqrt{1 - \frac{B(\rho, \sigma)^2}{4}}.
\end{equation}

Based on the Bures distance, we can calculate the total system evolution speed
\begin{equation} \label{utott}
u_\tot (t) = \lim_{\delta t \to 0} \frac{B(|\psi(t)\rangle, |\psi(t + \delta t)\rangle)}{\delta t},
\end{equation}
and the subsystem evolution speed
\begin{equation} \label{uAt}
u_A (t) = \lim_{\delta t \to 0} \frac{B(\rho_A (t), \rho_A (t + \delta t))}{\delta t}.
\end{equation}
These speeds satisfy the relation
\begin{equation}
u_A (t) \leq u_\tot (t) = \Delta H(t),
\end{equation}
with $\Delta H(t)$ being the variance of the Hamiltonian.

For two density matrices $\rho$ and $\sigma$, one could define the Schatten $n$-distance as
\begin{equation}
S_n(\rho,\sigma) = \Big[ \frac{1}{2} \text{tr} \big( |\rho-\sigma|^n \big) \Big]^{1/n},
\end{equation}
with the real number $n \geq 1$.
For $n=1$, it is just the trace distance.
We will use the Schatten 2-distance, also known as the Frobenius distance and the Hilbert-Schmidt distance, given by
\begin{equation}
S(\rho,\sigma) = \sqrt{\frac{1}{2} \text{tr}[(\rho-\sigma)^2]}.
\end{equation}
We will also use the normalized Schatten 2-distance \cite{Fagotti:2013jzu}, defined as
\begin{equation}
N(\rho,\sigma) = \sqrt{\frac{\text{tr}[(\rho-\sigma)^2]}{\text{tr}(\rho^2)+\text{tr}(\sigma^2)}}.
\end{equation}
For states in a finite-dimensional space, all the definitions of the distance are equivalent.
However, in the infinite-dimensional limit, the trace distance is preferred over the normalized Schatten 2-distance, in the sense that the trace distance can distinguish some states that the normalized Schatten 2-distance cannot \cite{Zhang:2020mjv}.
In the same sense, the normalized Schatten 2-distance is better than the Schatten 2-distance.
The advantage of the Schatten 2-distance and normalized Schatten 2-distance is that they are sometimes easier to calculate for large systems.
For a time-dependent state, we may define the different evolution speeds as
\begin{align}
s(t) & \equiv \lim_{\delta t \to 0} \frac{S(\rho(t),\rho(t+\delta t))}{\delta t}, \\
n(t) & \equiv \lim_{\delta t \to 0} \frac{N(\rho(t),\rho(t+\delta t))}{\delta t}.
\end{align}

For two density matrices $\rho$ and $\sigma$, one can define the relative entropy as
\begin{equation}
S(\rho\|\sigma) = \text{tr}(\rho\log\rho) - \text{tr}(\rho\log\sigma),
\end{equation}
which is also known as the Kullback-Leibler divergence.
The relative entropy is not a well-defined distance metric but can also characterize the difference between two states.
Furthermore, it is finite only when $\text{supp}(\rho) \subseteq \text{supp}(\sigma)$.
One should be cautious when using the relative entropy as a quantitative measure of the difference between two states.
The relative entropy bounds the trace distance according to Pinsker's inequality \cite{Watrous:2018rgz}
\begin{equation} \label{pinsker}
D(\rho,\sigma) \leq \sqrt{\frac{1}{2} S(\rho\|\sigma)}.
\end{equation}
This motivates us to define the ``{\it relative distance}''
\begin{equation} \label{relativedistance}
R(\rho\|\sigma) \equiv \sqrt{\frac{1}{2} S(\rho\|\sigma)},
\end{equation}
and the corresponding evolution speed
\begin{equation}
r(t) \equiv \lim_{{\delta t} \to 0} \frac{R(\rho(t)\|\rho(t+\delta t))}{\delta t}.
\end{equation}

\section{Transverse field Ising model}\label{supp:s3}

\begin{figure*}[t]
  \includegraphics[height=0.19\textwidth]{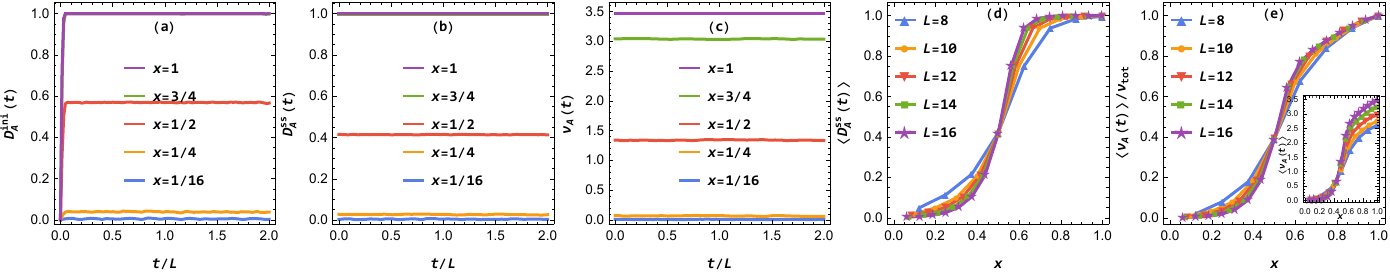}\\
  \caption{%
  {\bf Transverse field Ising model}: Subsystem trace distance and subsystem evolution speed with an initial random state.
  (a) Dynamics of the trace distance $D_A^\ini(t)$ between the RDM at time $t$ and the RDM at the initial time.
  (b) Dynamics of the trace distance $D_A^\ss(t)$ between the RDM at time $t$ and its corresponding relaxation state RDM.
  (c) Dynamics of the subsystem evolution speed $v_A(t)$.
  (d) Long-time average of $D_A^\ss(t)$ as a function of the ratio $x = L_A/L$.
  (e) Long-time average of $v_A(t)$ relative to the total system evolution speed $v_\tot$, plotted as a function of $x$. The inset shows the time-averaged subsystem evolution speed without normalization.
  In all panels, the transverse field is $h_z = \sqrt{2}$.
  For panels (a), (b), and (c), a total system size of $L = 16$ is used.
  The time averages for the distance and speed are taken over the interval $t \in [L, 2L]$.%
  }
  \label{FigureTFIMRandom}
\end{figure*}

We consider various initial states in the transverse field Ising model with the Hamiltonian
\begin{equation}
H(h_z) = -\frac{1}{2} \sum_{j=1}^{L} (\sigma_j^x \sigma_{j+1}^x + h_z\sigma_j^z).
\end{equation}
We first consider the initial random state.
The corresponding steady state is still the maximally mixed state $\r_\ss=1/2^L$.
The dynamics of subsystem trace distance and subsystem evolution speed in shown in Fig.~\ref{FigureTFIMRandom}.
It is the same as Fig.~\ref{FigureChaoticIsingRandom} in the main text for the initial random state in the chaotic Ising chain.
This is expected, as any unitary transformation of a random state results in a new random state as long as the unitary transformation is unrelated to the random state. The initial random state in other models are similar and we will not show in the  appendix.
\begin{figure*}[t]
  \includegraphics[height=0.19\textwidth]{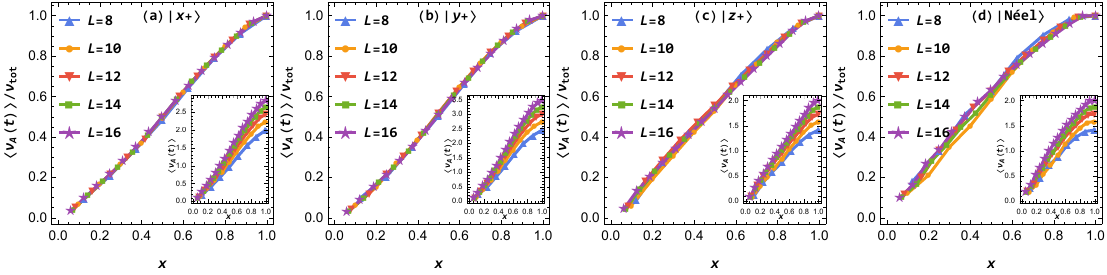}\\
  \caption{%
  {\bf Transverse field Ising model}: Long-time average of the subsystem evolution speed over the total system evolution speed, with the initial state being a product state: (a) $|x+\rangle$, (b) $|y+\rangle$, (c) $|z+\rangle$, and (d) $|\text{N\'{e}el}\rangle$.
  The insets show the subsystem evolution speed without normalization.
  In all panels, the transverse filed is $h_z = \sqrt{2}$.
  The time average is taken over the interval $t \in [L, 2L]$.%
  }
\label{FigureTFIMProduct}
\end{figure*}
Then we consider the example of the initial product states $%
|x+\rangle \equiv \bigotimes_{j=1}^{L} \frac{| \uparrow_j \rangle + | \downarrow_j \rangle}{\sqrt{2}}, ~
|y+\rangle \equiv \bigotimes_{j=1}^{L} \frac{| \uparrow_j \rangle + \ii | \downarrow_j \rangle}{\sqrt{2}}, ~
|z+\rangle \equiv \bigotimes_{j=1}^{L} | \uparrow_j \rangle, ~
|\text{N\'{e}el}\rangle \equiv |\uparrow\downarrow\uparrow\downarrow\cdots\rangle%
$, %
as shown in Fig.~\ref{FigureTFIMProduct}. There are no signs of relaxation for these states.

\begin{figure*}[t]
  \includegraphics[height=0.78\textwidth]{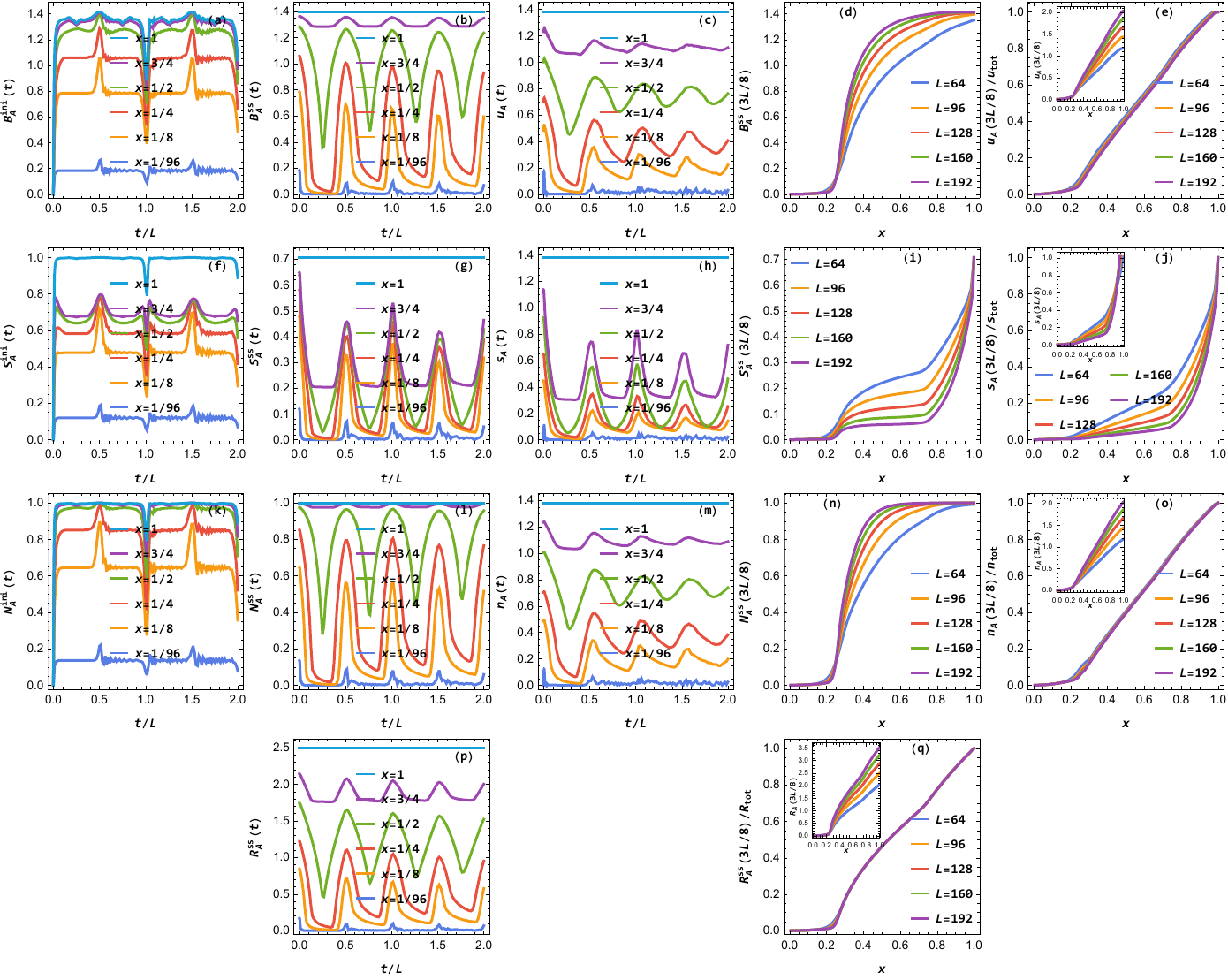}
  \caption{%
  {\bf Transverse field Ising model}: Subsystem distance and the corresponding subsystem evolution speed after a quantum quench $h_0 \to h_1$ with the distance defined as the Bures distance (the first row), the Schatten 2-distance (the second row), normalized Schatten 2-distance (the third row), and relative distance (the fourth row).
  Panels (a), (f) and (k): Dynamics of the subsystem distance between the time-dependent RDM and the initial RDM.
  Panels (b), (g), (l) and (p): Dynamics of the subsystem distance between the RDM and its corresponding GGE state RDM.
  Panels (c) and (h): Dynamics of the subsystem evolution speed.
  Panels (d), (i) and (n): Subsystem distance at a fixed time $t_*=\frac{3L}{8}$ as a function of the ratio $x = \frac{L_A}{L}$.
  Panel (q): Subsystem relative distance at a fixed time $t_*=\frac{3L}{8}$ over the total system relative distance. The inset indicates the subsystem relative distance without normalization.
  Panels (e), (j) and (o): Subsystem evolution speed at a fixed time $t_*=\frac{3L}{8}$ over the total system evolution speed. The insets indicate the subsystem evolution speed without normalization.
  We have used the numerical parameters $h_0=\sqrt{2}$ and $h_1=1$.
  In the first, second and third columns, we have used the total system size $L=96$.%
  }
  \label{FigureTFIMQuench}
\end{figure*}

Following the protocol of quantum quench in \cite{Calabrese:2005in,Calabrese:2006rx}, we can choose the initial state to be the ground state of the Hamiltonian $H(h_0)$ and then let it evolve under the Hamiltonian $H(h_1)$, where we denote the initial Hamiltonian parameter as $h_0$ and the post-quench Hamiltonian parameter as $h_1$. As all relevant RDMs are Gaussian states, and the fidelity between Gaussian states can be efficiently calculated following \cite{Banchi:2013uht}, we utilize the Bures distance (\ref{Brhosigma}) and the corresponding evolution speeds (\ref{utott}) and (\ref{uAt}) for the transverse field Ising model.
We present the subsystem Bures distance and the corresponding subsystem evolution speed in the first row of Fig.~\ref{FigureTFIMQuench}, where we have used $h_0=\sqrt{2}$ and $h_1=1$. As shown in panels (a) and (b), the states evolve with a period $L$ after the quantum quench $h_0 \to h_1$, and subsystems smaller than half of the total system thermalize to the RDMs of the generalized Gibbs ensemble (GGE) with a period of $\frac{L}{2}$. For a subsystem of size $0<L_A<\frac{L}{2}$ within the period $t\in[0,\frac{L}{2}]$, thermalization occurs in the time range $\frac{L_A}{2}<t<\frac{L-L_A}{2}$. In other words, at some time $0<t_*<\frac{L}{2}$, thermalization occurs for a subsystem of size $0<L_A<\min(2t_*,L-2t_*)$.
In panel (c), the subsystem evolution speed $u_A(t)$ exhibits similar dynamics to the subsystem Bures distance $B_A^\ss(t)$. In panels (d) and (e), we fix the time at $t_*=\frac{3L}{8}$, where thermalization occurs for subsystems with $0<x<x_*=\frac{1}{4}$. All these results are consistent with the thermalization and revival phenomena described in \cite{Cardy:2014rqa} and the quasiparticle picture presented in \cite{Calabrese:2005in,Alba:2017hlc} for integrable models.

The second and third rows of Fig.~\ref{FigureTFIMQuench} display results for the subsystem Schatten 2-distance and its corresponding evolution speed, and the normalized Schatten 2-distance and its corresponding evolution speed, post-quantum quench in the finite size TFIM. This indicates that the normalized Schatten 2-distance and its evolution speed are effective indicators of equilibration, whereas the unnormalized versions are not. It is noteworthy that, despite not being proper equilibration indicators, the subsystem Schatten 2-distance and its evolution speed exhibit similar behaviors.

We present the subsystem relative distance (\ref{relativedistance}) following the quantum quench evolution in the fourth row of Fig.~\ref{FigureTFIMQuench}. This demonstrates that, when well-defined, the relative distance serves as a good relaxation indicator. We chose not to include other relative distances and their associated evolution speeds in this analysis, as they are not well-defined.

\section{Indications for fluctuations of operator expectation value}\label{supp:s4}

The ETH proposes an ansatz regarding how the matrix elements of observables behave across the eigenstates of a Hamiltonian. Within this framework, it has been shown that the long-time average of any local observable converges to its expected value as predicted by statistical mechanics, provided that the energy fluctuations within the diagonal ensemble remain small. Furthermore, leveraging the ETH ansatz enables the calculation of the long-term average of the temporal fluctuations in the expectation value of an observable. These fluctuations are found to be exponentially small relative to the system size \cite{DAlessio:2015qtq,Borgonovi:2016hnp}. In this context, we demonstrate that the exponential suppression of these fluctuations may be attributed to the exponential decrease in subsystem evolution speeds as the system size increases.

We consider the subsystem evolution speed given by
\begin{equation}
v_A(t) = \f12 \Big\| \f{\dd \r_A(t)}{\dd t} \Big\|_1.
\end{equation}
For a time interval $t \in [t_0, t_0 + T]$ where $T$ is at most polynomial in the system size $L$, we assume that the subsystem evolution speed is exponentially small, as
\begin{equation}
v_A(t) \sim e^{-\alpha L},
\end{equation}
where $\a$ is a constant independent of $t$ or $L$.
The RDM can be expressed as
\begin{equation}
\rho_A(t) = \rho_A(t_0) + \int_{t_0}^{t} \!\!\dd t^\prime \frac{\dd\rho_A(t^\prime)}{\dd t^\prime}.
\end{equation}
The time average RDM in the period $t\in[t_0,t_0+T]$ is then
\begin{align}
%\begin{split}
\overline{\r_A} &= \frac{1}{T} \int_{t_0}^{t_0+T} dt \rho_A(t)
                \notag\\& = \rho_A(t_0)+ \frac{1}{T} \int_{t_0}^{t_0+T} \!\!\dd t \int_{t_0}^{t} \!\!\dd t^\prime \frac{\dd\rho_A(t^\prime)}{\dd t^\prime}.
%\end{split}
\end{align}
Given the exponentially small subsystem evolution speed, the subsystem trace distance is also exponentially small. Using the triangle inequality of the trace norm, the distance between the RDM and its time average is
%\begin{align} \label{DrAtoverlinerA}
%D( \r_A(t), \overline{\r_A} ) & = \f12 \Big\| \int_{t_0}^t \dd t' \f{\dd \r_A(t')}{\dd t'} - \f{1}{T} \int_{t_0}^{t_0+T} \dd t \int_{t_0}^t \dd t' \f{\dd \r_A(t')}{\dd t'} \Big\|_1 \leq \f12  \int_{t_0}^t \dd t' \Big\| \f{\dd \r_A(t')}{\dd t'} \Big\|_1
%+ \f{1}{2T} \int_{t_0}^{t_0+T} \dd t \int_{t_0}^t \dd t' \Big\| \f{\dd \r_A(t')}{\dd t'} \Big\|_1 \\
%&  = \int_{t_0}^t \dd t' v_A(t') + \f{1}{T} \int_{t_0}^{t_0+T} \dd t \int_{t_0}^t \dd t' v_A(t') \sim \ep^{-\a L}.
%\end{align}
%\begin{widetext}
\begin{align} \label{DrAtoverlinerA}
D( \r_A(t), \overline{\r_A} ) &\leq \int_{t_0}^t \!\!\dd t' v_A(t') \notag \\&+ \f{1}{T} \int_{t_0}^{t_0+T} \!\!\dd t \int_{t_0}^t \dd t' v_A(t')
                              \sim \ep^{-\a L}.
\end{align}
%\end{widetext}
Indeed, the trace distance between the RDMs at any two different times $t_0 \leq t_1 < t_2 \leq t_0 + T$ is exponentially small
\begin{equation}
%\begin{split}
D(\rho_A(t_1), \rho_A(t_2))  %&= \frac{1}{2} \left| \left| \int_{t_1}^{t_2} dt^\prime \frac{d\rho_A(t^\prime)}{dt^\prime} \right|\right|_1 \\&
\leq \left|\left| \int_{t_1}^{t_2} \!\!\dd t^\prime v_A(t^\prime) \right| \right|_1 \sim e^{-\alpha L}.
%\end{split}
\end{equation}
For an operator $\cO$ defined on subsystem $A$, the time average of the expectation value is
\begin{align}
\overline{\langle \cO \rangle} &= \frac{1}{T} \int_{t_0}^{t_0+T} \!\!\dd t \langle \psi(t) |\cO| \psi(t) \rangle \notag \\
                               &= \frac{1}{T}\int_{t_0}^{t_0+T} \!\!\dd t \tr_A( \r_A(t) \cO ) \notag = \tr_A( \overline{\r_A} \cO ).
\end{align}
From the H\"older inequality and the exponentially small trace distance, it follows that
\begin{equation}
\big|\langle \psi(t) |\cO| \psi(t) \rangle - \overline{\langle \cO \rangle}\big| \leq 2s_\cO D(\rho_A(t), \overline{\r_A}) \sim e^{-\alpha L},
\end{equation}
where $s_\cO$ is the largest singular value of the operator $\cO$.
This indicates exponentially small time fluctuations of the operator's expectation value
\begin{equation}
\f{1}{T} \int_{t_0}^{t_0+T} \!\!\dd t \big| \lag\psi(t)|\cO|\psi(t)\rag - \overline{\lag \cO \rag} \big|^2 \sim \ep^{-2\a L}.
\end{equation}

\section{Small subsystem evolution speed from small subsystem trace distance}\label{supp:s5}

The exponentially small subsystem trace distance indicates that the subsystem evolution speed is similarly negligible. Specifically, within a given time interval $t \in [t_0, t_0 + T]$, if the trace distance between the subsystem state $\rho_A(t)$ and the steady state $\rho_{A,\ss}$ satisfies
\begin{equation}
D(\rho_A(t), \rho_{A,\ss}) \sim e^{-\alpha L},
\end{equation}
for some constant $\alpha$, then we can use the triangle inequality of the trace distance to show that
\begin{align}
D(\rho_A(t), \rho_A(t + \delta t)) &\leq D(\rho_A(t), \rho_{A,\ss}) \notag \\&+ D(\rho_A(t + \delta t), \rho_{A,\ss}) \sim e^{-\alpha L}.
\end{align}

By choosing $\delta t$ such that $e^{-\alpha L} \ll \delta t \ll 1$, we can infer that the evolution speed $v_A(t)$ is also exponentially small, given by
\begin{equation}
v_A(t) = \lim_{\delta t \to 0} \frac{D(\rho_A(t), \rho_A(t + \delta t))}{\delta t}\sim e^{-\alpha L}.
\end{equation}

This result has important consequences for the time fluctuation of operator expectation values within subsystem $A$. For any operator $\cO$ defined within subsystem $A$, the H\"older inequality bounds the difference between the expectation value of $\cO$ in the state $\psi(t)$, $\langle \psi(t)|\cO|\psi(t)\rangle=\tr_A(\r_A(t)\cO)$, and its steady state expectation value $\langle \cO \rangle_{\ss}=\tr(\r_\ss\cO)=\tr_A(\r_{A,\ss}\cO)$. Specifically,
\begin{equation}
\big|\langle \psi(t)|\cO|\psi(t)\rangle - \langle \cO \rangle_{\ss}\big| \leq 2s_\cO D(\rho_A(t), \rho_{A,\ss}) \sim e^{-\alpha L},
\end{equation}
where $s_\cO$ is the largest singular value of the operator $\cO$. Consequently, the time fluctuation of the operator expectation value is also exponentially small, as evidenced by the scaling
\begin{equation}
\f{1}{T}\int_{t_0}^{t_0+T} \!\!\dd t \big| \langle \psi(t)|\cO|\psi(t)\rangle - \langle \cO \rangle_{\ss} \big|^2 \sim e^{-2\alpha L}.
\end{equation}

\end{document}